\documentclass[intlimits,twoside,a4paper]{article}
\usepackage{amsmath,amssymb}
\usepackage{graphicx}
\usepackage{wrapfig}
\usepackage[T2A]{fontenc}
\usepackage[cp1251]{inputenc}
\usepackage{cmpj}

\issue{2011}{14}{3}{33604}

\doinumber{10.5488/CMP.14.33604}

\title[Changes in the structure of tethered chains]%
{Changes in the structure of tethered chain molecules as predicted by density functional approach\thanks{Dedicated to Professor Yu. Kalyuzhnyi on the occasion of his 60th birthday.}
}
\author[M. Bor\'owko {\sl et. al}]{M. Bor\'owko\refaddr{label1}, A. Patrykiejew\refaddr{label1},
O. Pizio\refaddr{label2}, S. Soko\l owski\refaddr{label1} 
}

\addresses{
\addr{label1} Department for the Modelling of Physico-Chemical Processes, \\Maria
Curie-Sk\l odowska University, 20031 Lublin, Poland
\addr{label2} Instituto de Qu\'{i}mica, Universidad Nacional Autonoma de M\'{e}xico,
Circuito Exterior, \\Ciudad Universitaria, 04360, M\'{e}xico, D.F., M\'{e}xico
}

\authorcopyright{M.~Bor\'owko, A.~Patrykiejew, O.~Pizio, S. Soko\l owski, 2011}

\date{Received April 1, 2011, in final form June 16, 2011}

\begin{document}

\maketitle

\begin{abstract}
We use a version of the density functional theory to study the changes in the height of the tethered
layer of chains built of jointed spherical segments with the change of the length and surface density of chains.
For the model in which the interactions between segments and solvent molecules are the same as
between solvent molecules we have discovered two effects that have not been observed in
previous studies. Under certain conditions and for low surface concentrations of the chains, the height
of the pinned layer may attain a minimum. Moreover, for some systems we observe that when the
temperature increases, the height of the layer of chains may decrease.
\pacs 68.47.Mn, 61.25.H-, 68.47.Pe, 82.35.Gh
\end{abstract}

\section{Introduction}
The materials modified with tethered chains play  an important role in several technological fields.
 With the proper choice of  the grafting surface, the length of chains and
the grafting density, one can obtain novel products exhibiting desired behavior. The materials modified in
such a way  are utilized in a wide range of applications including chromatography,
adhesion, lubrication and colloidal stabilization~\cite{3,5,6}.  
Recent studies  have focused on  the
development of  ``smart'' materials,  whose properties can be readily altered by applied stimuli. They can find
 applications in drug delivery, as chemical sensors and in controlling the nanoparticle transport~\cite{9,10,12,14}.
 Furthermore, applications in
nanotechnology include end grafted polymers on nanopatterned surfaces, on
nanoparticles,  or on carbon nanotubes~\cite{15,16,18,19,20}.

Much effort has been directed toward theoretical description of
grafted polymer layers. The studies included scaling theories~\cite{21,22,22a},
classical self-consistent field methods~\cite{23,24}, single chain mean-field methods~\cite{25},
density functional theories~\cite{25,26,27,28,29}  and computer simulations ~\cite{30,31,32,32a,33,34,35,36,37,38,38a,38b}.
In our opinion, particularly important for further theoretical studies of systems
involving tethered chains is the development of density functional theory (DFT)~\cite{27,28,39,40,41,42,43,44,45,46}
that is based on the approach proposed by Yu and Wu~\cite{47}.
 In general, this type of theory represents a powerful tool for the study of adsorption and
phase behavior of confined systems.  Within this approach,
theoretical description of ``pinning'' the  terminating segments
of chain molecules to the walls can be realized by using the ideas of the previously developed
methods of bonding the fluid species to the walls in an associative treatment~\cite{48,49}.

Recent theoretical works based on the DFT ~\cite{27,28,39,40,41,42,43,44,45,46} have
mainly concentrated on the description of adsorption of fluids on modified surfaces,
while the problem of the changes in the layer of pinned chains itself has attracted less
attention.  Of course, the accumulation of fluid molecules near the adsorbing surface
induces changes in the tethered layer, especially when the adsorbed fluid undergoes prewetting of
layering transitions~\cite{44,46}. However, no systematic density functional  study of changes in
the structure of tethered chains has been undertaken so far.

The problem of a change of the configuration of tethered chains with
their length, surface density and the quality of a solvent has been considered in a great
number of publications. One of the first theoretical works devoted to that problem were
the publications by Alexander and de Gennes~\cite{21,22}. However, the references to
numerous more recent  works can be found in the review article of Descas et al.~\cite{50}
and in the PhD thesis by Chremos~\cite{51}. We know that in a good solvent,
 for not too high surface densities of tethered chains and when the free segments (i.e. all but the tethered segment)
are not too strongly attracted by the surface, the chains assume the so-called mushroom configuration. With
an increase of the surface density,  the overlapping mushroom configuration is observed. This
configuration precedes the brush regime that occurs at high surface concentrations.
Despite the great number of experimental, theoretical and computational
investigations on the behavior of end anchored chains, the quantitative
determination of the parameter space appropriate for detailed studies  of
 transitions between possible configurations of tethered chains is still vague.
 In addition, the predictions of theories do not always
coincide with the results of simulations~\cite{38,52}.  Moreover,  in our opinion
insufficient attention has been paid to some problems, such
as the effect of competition between strong adsorption of fluid
molecules and adsorption of free segments of a chain on the structure of the chains and to
the effect of the architecture of branched polymers on the structure of the formed layer~\cite{53}.

The  description of systems comprising tethered chains
at a microscopic level requires the knowledge
of numerous parameters. One has to know not only
the surface density and the length of the chains, ``quality''  (and density) of the solvent and the temperature,
but also details about the architecture of the chains, parameters describing the interactions
between all the components themselves and between the components and  the surface, and, in some cases,
the topography of the end anchored chains on the surface. A huge number of
parameters of the model delimits the possibility of an effective scanning of the space of
the model parameters for a precise evaluation of the regions of different configurations of chains and
the transitions between them by long lasting computer simulations. Such studies, however,
can be carried out by using theoretical approaches based on microscopic models.
Therefore, the aim of the present work is to employ the DFT used in our previously
works~\cite{27,28,43,44,45,46}  to investigate how the thickness of the pinned layer changes
with their length and with their surface density.  Our principal purpose is to check
whether the theory employed by us is capable of reproducing the results of coarse-grained approaches~\cite{21,22,22a}. We are also looking for some new effects that have not been observed in
previous studies.

\section{Theory}

Following our previous works~\cite{27,28,43,46}
we consider a fluid of spherical molecules of diameter $\sigma_{\mathrm{f}}$
in contact with a modified substrate. The surface is covered by
the film of preadsorbed chain
molecules.  All the chains are represented by tangentially
jointed $M$ spherical beads of the same diameter, $\sigma_{\mathrm{c}}$. The chain connectivity is enforced
by the bonding potential between nearest-neighbor segments, $V_{\mathrm{b}}$. This potential satisfies
the equation~\cite{47}
\begin{equation}
\exp[-\beta V_{\mathrm{b}} (\mathbf{R})] =\prod_{i=1}^{M-1}
\delta(|r_i - r_{i+1} | - \sigma_{\mathrm{c}} )/4\pi\sigma_{\mathrm{c}}
\end{equation}
where $\mathbf{R} = (\mathbf{r}_1 ,\mathbf{r}_2 , \dots, \mathbf{r}_M )$  is a vector specifying segment positions
of consecutive segments.
Each polymer molecule of the chemically bonded phase contains
one surface-binding segment located at its end (indexed as ``1'') that
interacts with the wall via the potential
\begin{equation}
\exp[-\beta v_{\mathrm{s}1} (z)] = C\delta(z-\sigma_{\mathrm{c}}/2)
\end{equation}
where $z$ is a distance from the surface and $C$ is a constant.
This potential implies that the surface-binding segments always lie at the distance
$x=\sigma_{\mathrm{c}}/2$ from the
surface. The remaining segments of the grafted molecules ($j = 2, 3,\dots, M$) are ``neutral'' with respect to the surface,
i.e., they interact with the surface via a hard-wall potential
\begin{equation}
 v_{\mathrm{s}j}(z)=\left\{
\begin{array}{ll}
 \infty, & z<\sigma_{\mathrm{c}}/2, \\
0, & z\geqslant  \sigma_{\mathrm{c}}/2.
\end{array}
\right.
\end{equation}
The fluid molecules, however, interact with the surface via the Lennard-Jones (9,3)
potential of the form.
\begin{equation}
 v(z)=\varepsilon_{\mathrm{fs}}\left[ (z/z_0)^9-(z/z_0)^3\right],
\end{equation}
with $z_0=\sigma_{\mathrm{f}}/2$.

We assume the Lennard-Jones (12,6) type interactions between all
segments, between the segments and fluid particles and between the fluid particles
\begin{equation}
 u_{ij}(r)=\varepsilon_{ij}\left[ (r/\sigma_{ij})^{12}-(r/\sigma_{ij})^6\right],
\end{equation}
where $\sigma_{ij}=0.5(\sigma_i +\sigma_j),$ $i,j=c,f$.
Employing the perturbative treatment we split these interactions into the
repulsive (reference) and attractive (perturbation)
parts according to the Weeks-Chandler-Anderson scheme~\cite{54}
$u_{ij}(r)=u_{ij,\mathrm{ref}}(r)+u_{ij,\mathrm{pert}}(r)$, where
\begin{equation}
 u_{ij,\mathrm{pert}}(r)=\left\{
\begin{array}{ll}
 -\varepsilon_{ij}, & r<2^{1/6}\sigma_{ij}\,,\\
u_{ij}(r), & r\geqslant  2^{1/5}\sigma_{ij}\,.
\end{array}
\right.
\end{equation}

In order to proceed, let us introduce the notation, $\rho^{(c)}(\mathbf{R})$ and $\rho(z)$,
for the density distribution of chains and of spherical species, respectively.
However, the theory is constructed in terms of
the density of particular segments of chains, $\rho_{\mathrm{s}j}(z)$,  and  the
total segment density of chains, $\rho_{\mathrm{s}}(z)$. These densities are introduced via commonly used
relations, see the original development in~\cite{27,28,43,44,45,46}
\begin{equation}
 \rho_{\mathrm{s}}(\mathbf{r})=\sum_{j=1}^M \rho_{\mathrm{s}j}(\mathbf{r})=\sum_{j=1}^M
 \int \rd\mathbf{R}\delta(\mathbf{r}_j-\mathbf{r})\rho^{(c)}(\mathbf{R}).
\end{equation}
In the system under study all the local densities are the functions of the distance from the
surface only.

The system is studied in a grand canonical ensemble with the constraint of constancy of the
number of chain molecules, i.e.
\begin{equation}
\rho_{\mathrm{c}}=\int \rd z \rho_{\mathrm{s}j}(z)
\label{eq:con}
\end{equation}
where $\rho_{\mathrm{c}}$ is the number of chain molecules per area of the surface, i.e. it is the
surface density of the chains. Of course, the integral in equation~(\ref{eq:con}) does not depend on the
segment index $j$.

The thermodynamic potential
appropriate to the description of the system is
\begin{equation}
 Y=F[\rho_{\mathrm{s}}(z),\rho(z)]+\int \rd z \rho(z)[v(z)-\mu]+\sum_{j=1}^M\int \rd z \rho_{\mathrm{s}j}v_{\mathrm{s}j}(z),
\label{eq:t}
\end{equation}
where $F[\rho_{\mathrm{s}}(z),\rho(z)]$ is the Helmholtz free energy functional and $\mu$ is the chemical potential
of the reference system fluid. The expression for $F[\rho_{\mathrm{s}}(z),\rho(z)]$  is obtained from the theory
described in our earlier works~\cite{27,28,43,44,45,46} and in the original papers by Yu and Wu~\cite{47}.
 The density profile of fluid molecules and the segment
density profiles are obtained by minimizing the functional~(\ref{eq:t}) under the constraint (\ref{eq:con}).
For the sake of brevity we do not present the resulting density profile equations. They can be found
in  the above cited publications.
We should only note  that the reference system comprises only fluid species. The chemical potential of
the fluid, $\mu$  is thus given by
\begin{equation}
 \mu=kT\ln\rho_{\mathrm{b}} +\mu_{\mathrm{hs}}+\rho_{\mathrm{b}}\int u_{\mathrm{ff},\mathrm{pert}}(r)\rd\mathbf{r},
\end{equation}
 where $\rho_{\mathrm{b}}$ is the reference system density and  $\mu_{\mathrm{hs}}$ is the chemical potential of hard-spheres of
diameter $\sigma_{\mathrm{f}}$, resulting from the Carnahan-Starling equation of state.

The system under study is described by several parameters. In order to reduce the number of
parameters to a minimum, we assume that $\sigma_{\mathrm{f}}=\sigma_{\mathrm{c}}=\sigma_{\mathrm{cf}}\equiv\sigma$ and that
$\varepsilon_{\mathrm{ff}}=\varepsilon_{\mathrm{cf}}\equiv\varepsilon$. The Lennard-Jones (12,6) potential parameter, characterizing
the interactions between segments of the chain, $\varepsilon_{\mathrm{cc}}$,  as well as
the fluid-solid energy parameter are expressed in the units of $\varepsilon$, $\varepsilon_{\mathrm{cc}}^*=\varepsilon_{\mathrm{cc}}/\varepsilon$
and  $\varepsilon_{\mathrm{fs}}^*=\varepsilon_{\mathrm{fs}}/\varepsilon$.  The reduced densities are defined as usual,
$\rho_{\mathrm{b}}^*=\rho_{\mathrm{b}}^{}\sigma^3$, $\rho^*(z)=\rho(z)\sigma^3$,  $\rho^*_{\mathrm{s}j}(z)=\rho_{\mathrm{s}j}(z)\sigma^3$,
$\rho^*_{\mathrm{s}}(z)=\rho_{\mathrm{s}}(z)\sigma^3$ and $\rho_{\mathrm{c}}^*=\rho_{\mathrm{c}}\sigma^2$. Moreover, we introduce
the reduced temperature as $T^*=kT/\varepsilon$.

Almost all the calculations (unless otherwise explicitly stated) have been carried out for the fluid
bulk density $\rho_{\mathrm{b}}^*=0.66$. This density nearly coincides with the liquid density on the
liquid-vapor coexistence curve at $T^*=1$.

\section{Results and discussion}

Before discussing the results we should emphasize that
it is unclear how to determine  the brush height in a precise meaning from the
density profiles obtained via DFT or from simulations. It has been a
widespread practice to define the brush
height, $\langle h \rangle$, from the first moment of the total segment density profile~\cite{32a,55,56}
\begin{equation}
 \langle h \rangle=a\frac{\int \rd z z \rho_{\mathrm{s}}(z)} {\int \rd z \rho_{\mathrm{s}}(z)}\,.
\end{equation}
Usually,  (cf.~\cite{32a,55}) the coefficient $a$ was   set to be $a=8/3$, but also $a=2$ was used (see, e.g.~\cite{56})
and the latter value has been adopted in our calculations. Moreover, usually the height of the brush,
as well as the surface density of tethered chains were scaled by the radius of gyration (or by some
powers of the radius of gyration) of the chain molecules, cf.~\cite{32a,38,52,55,56}.  Unfortunately, the value of the radius
of gyration cannot be self-consistently obtained within the framework of the considered approach. Therefore,
in our calculations we use the reduced units as defined at the end of the previous section.

\subsection{Change of the brush height with the length of chains}

\begin{figure}[ht]
\hspace{3mm}
\includegraphics[width=0.45\textwidth]{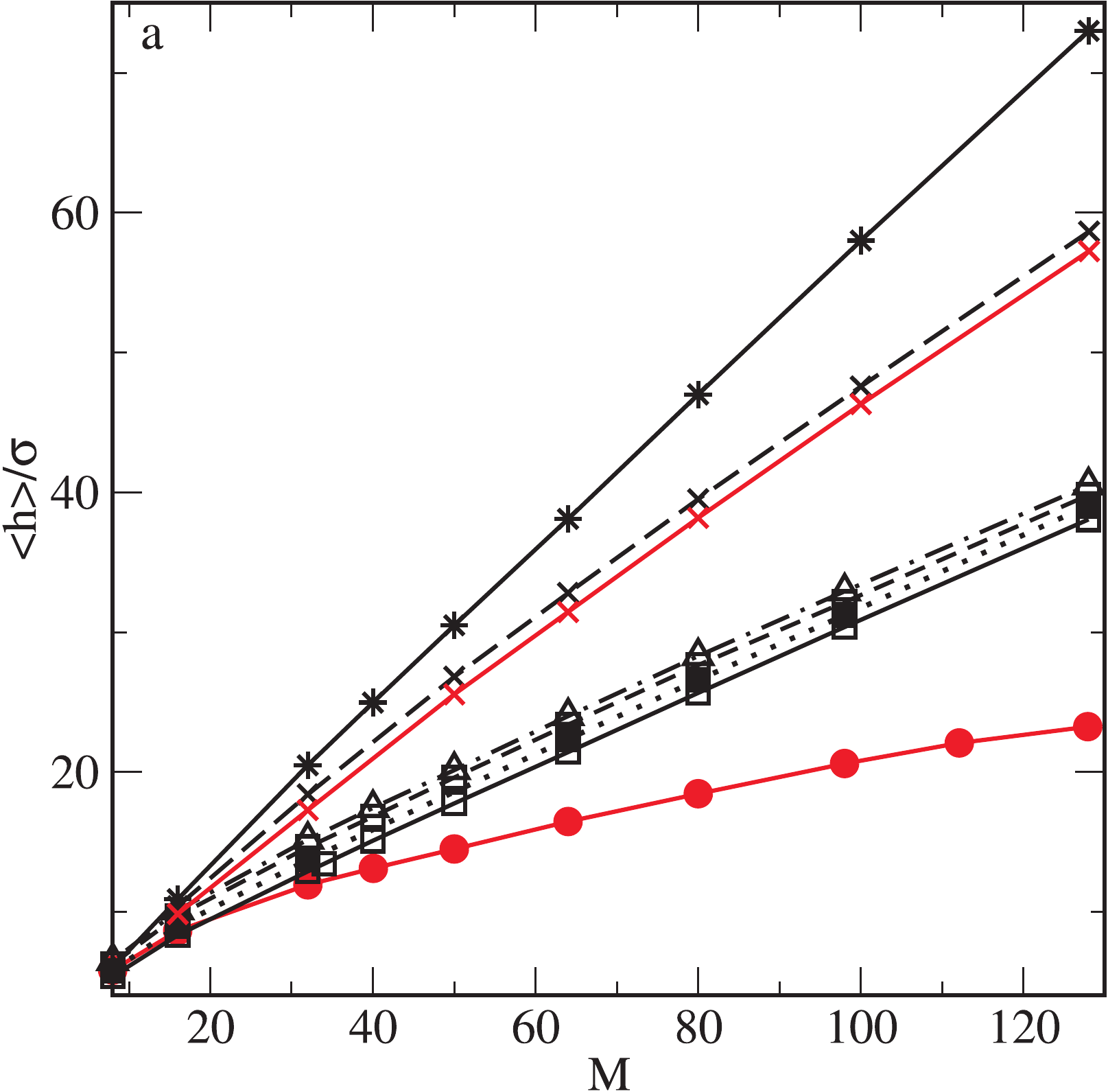}%
\hfill%
\includegraphics[width=0.45\textwidth]{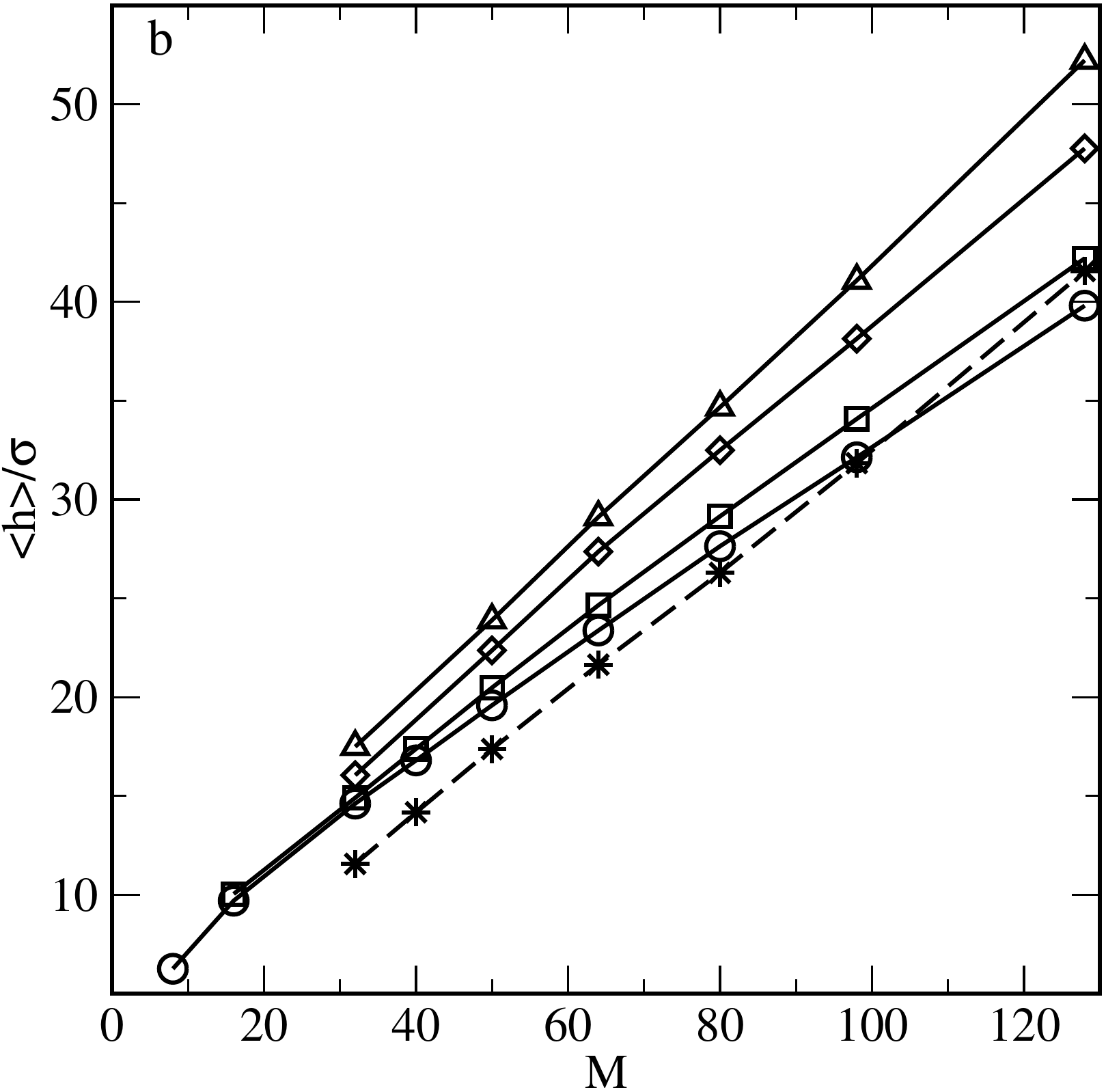}%
\hspace{3mm}
\\%
\parbox[t]{0.48\textwidth}{%
\centerline{(a)}%
}%
\hfill%
\parbox[t]{0.48\textwidth}{%
\centerline{(b)}%
}%
\caption{\looseness=1The dependence of  $\langle h \rangle/\sigma$  on $M$. In part (a)  the nomenclature of the lines decorated with different symbols
is as follows. Solid lines have been evaluated for $\varepsilon_{\mathrm{fs}}^*=1$ and  for the chains
built of hard-spheres, $\varepsilon_{\mathrm{cc}}^*=0$  (stars);  $\varepsilon_{\mathrm{cc}}^*=0.6$ (crosses)
$\varepsilon_{\mathrm{cc}}^*=1$ (open squares).
Dotted line with filled squares is for $\varepsilon_{\mathrm{fs}}^*=5$ and $\varepsilon_{\mathrm{cc}}^*=1$.
Dashed lines are for  $\varepsilon_{\mathrm{fs}}^*=12$ and
for   $\varepsilon_{\mathrm{cc}}^*=0.6$  (crosses)  and $\varepsilon_{\mathrm{cc}}^*=1$
(open squares), while dash-dotted line with triangles is for  $\varepsilon_{\mathrm{fs}}^*=18$ and $\varepsilon_{\mathrm{cc}}^*=1$.
The values of the remaining parameters are: $T^*=1$ and  $\rho_{\mathrm{c}}^*=0.1$.
Additionally, solid line with filled circles has been obtained for  $\varepsilon_{\mathrm{fs}}^*=1$ and $\varepsilon_{\mathrm{cc}}^*=1$,
$T^*=1$ and $\rho_{\mathrm{c}}^*=0.02$.
In part (b) we show the effect of surface density $\rho_{\mathrm{c}}^*$  and of the temperature on the evaluated curves
 $\langle h \rangle/\sigma=f(M)$. Solid  lines are at $T^*=1$  and  $\rho_{\mathrm{c}}^*=0.1$ (open circles),  $\rho_{\mathrm{c}}^*=0.12$ (open squares),
 $\rho_{\mathrm{c}}^*=0.15$ (open diamonds) and  $\rho_{\mathrm{c}}^*=0.20$ (open triangles). The dashed line with
stars is for  $\rho_{\mathrm{c}}^*=0.1$ and at $T^*=4$. The values of the remaining parameters are  $\varepsilon_{\mathrm{fs}}^*=12$ and
$\varepsilon_{\mathrm{cc}}^*=1$.}
\protect
\label{fig:1}
\end{figure}
We begin with the discussion of the
changes of the brush height with the length of the tethered chains. In figure~\ref{fig:1}
we show the plots of $\langle h  \rangle/\sigma$ versus $M$ for a number of systems characterized by
different values of the intermolecular potentials and
obtained for different surface grafting densities.
In part (a) the solid lines were obtained assuming weak interactions between fluid molecules and solid surface,
$\varepsilon_{\mathrm{fs}}^*=\varepsilon_{\mathrm{fs}}/\varepsilon=1$. The remaining lines
were evaluated for solid surfaces  attracting the fluid molecules more strongly,
see the figure caption.  The surface density of grafted chains is $\rho_{\mathrm{c}}^*=0.1$
for all, but the one curve, for which $\rho_{\mathrm{c}}^*=0.02$
(the solid line decorated with solid circles).
From first glance, for $M>10$,  almost all the curves plotted here
are well approximated by straight lines (an exception is the
solid line decorated with solid circles, calculated for $\rho_{\mathrm{c}}^*=0.02$).
However, much better approximations
are obtained having assumed a power-law dependence,
cf.~\cite{57}
\begin{equation}
\langle h  \rangle\propto M^\alpha.
\label{eq:pl1}
\end{equation}
Even for the segment-segment interactions of a hard-sphere
type (solid line decorated with stars) the approximation of numerical results by a straight
line of the form $\langle h  \rangle=a+bM$ yields  the correlation coefficient $R_{\mathrm{c}}$ of the order of
$R_{\mathrm{c}}\approx 0.990$, while  the use of equation~(\ref{eq:pl1}) gives $R_{\mathrm{c}}\approx 0.9999$ for
$\alpha=0.9134$.

When  the attractive interactions between the segments are turned on, the exponent $\alpha$
decreases. For $\varepsilon_{\mathrm{cc}}=0.6\varepsilon$ we obtain   $\alpha=0.850$
(solid line with crosses; the correlation coefficient in this case is again of the order of 0.9999),
while for  $\varepsilon_{\mathrm{cc}}=\varepsilon$  we have  $\alpha=0.738$ (solid line with open squares; the
correlation coefficient is now slightly lower, $R_{\mathrm{c}}\approx 0.999$). An increase  of the
fluid-solid energy decreases slightly the value of $\alpha$.
 For example, when  $\varepsilon_{\mathrm{cc}}=0.6\varepsilon$
and $\varepsilon_{\mathrm{fs}}^*=12$ (dashed line with crosses)
 we have $\alpha=0.834$, instead of $\alpha=0.850$ for $\varepsilon_{\mathrm{fs}}^*=1$
(solid line with crosses).
Similarly, for  $\varepsilon_{\mathrm{cc}}=\varepsilon$ and for $\varepsilon_{\mathrm{fs}}^*=18$
(dash-dotted line with triangles)
 we obtain $\alpha=0.720$. The last value is lower than that obtained for  $\varepsilon_{\mathrm{cc}}=\varepsilon$  and
$\varepsilon_{\mathrm{fs}}^*=1$.  The effect of the parameter $\varepsilon_{\mathrm{fs}}^*$  on
the exponent $\alpha$ is much weaker that of the parameter $\varepsilon_{\mathrm{cc}}$.

The solid line decorated with filled circles that has been evaluated for
much lower surface density of tethered chains  deviates very much from all the remaining results.
In this case the exponent of equation~(\ref{eq:pl1}) is much lower and equals $\alpha\approx 0.49$.
This indicates that the surface density of tethered chains
greatly effects the scaling relationship (\ref{eq:pl1}). To investigate this problem, we
calculated the values of $\langle h  \rangle/\sigma$ for different values of $\rho_{\mathrm{c}}^*$, keeping the values of the following
parameters constant:  $\varepsilon_{\mathrm{cc}}=\varepsilon$ and $\varepsilon_{\mathrm{fs}}^*=12$, see figure~\ref{fig:1}~(b).
 Solid lines displayed here
correspond to $T^*=1$, while the dashed line is for $T^*=4$. For $\rho_{\mathrm{c}}^*=0.1$
(solid line with open circles) we found that  $\alpha=0.729$, while for $\rho_{\mathrm{c}}^*=0.2$ (solid linear with triangles)~-- $\alpha=0.828$.  Therefore, we conclude that an increase of the surface density
of tethered chains increases the value of $\alpha$. This is a rather obvious result because in the case of
higher values of $\rho_{\mathrm{c}}^*$ the chains are more stretched and there is ``less room'' for lateral motions
of the segments which could possibly lead to ``coiling'' the chains.  However, the effect of the temperature on
the exponent $\alpha$ is not so obvious. The dashed line was obtained at $T^*=4$ for $\rho_{\mathrm{c}}^*=0.1$.
Now, $\alpha\approx 0.90$. For lower values of $M$ the height of the brush at $T^*=4$ is lower than
at $T^*=1$, but for longer chains the situation is reverse.  One can argue here that the average brush height,
$\langle h  \rangle$, should be
related to the end-to-end distance, $R_{1-N}$,
 of a single polymer chain, at least at low grafting density. Unfortunately, the
density functional approach used in this work does not allow for the evaluation of the end-to-end distance,
and therefore we cannot carry out any test for a relation between $\langle h \rangle $ and $R_{1-N}$ using
self-consistent data in the framework of the applied theory.

The configuration assumed by the chains
depends on entropic as well as on energetic contributions to the free energy. The chains
maximize their configurational entropy by adopting a random-walk-like configurations. However, depending on the energies of
segment-segment and segment-fluid attractions, they tend to maximize the contact between
relevant species. Moreover, the attraction between fluid molecules  and the solid leads to accumulation of
the fluid molecules in the vicinity of the solid surface.  At $T^*=4$ the Boltzmann factor
of the fluid-solid interactions is much lower than at $T^*=1$. Therefore, the accumulation of
fluid molecules and thus the effect of ``superseding'' the segments from the region close to the wall is
smaller at $T^*=4$. The adsorption of fluid molecules on the brush structure plays a more
significant role for shorter chains. For longer chains, however, the entropic effects prevail
and at $T^*=4$ the brush becomes more stretched than at $T^*=1$. We return to the problem of
 temperature dependence of the brush height hereinbelow.

\begin{figure}[ht]
\includegraphics[width=0.48\textwidth]{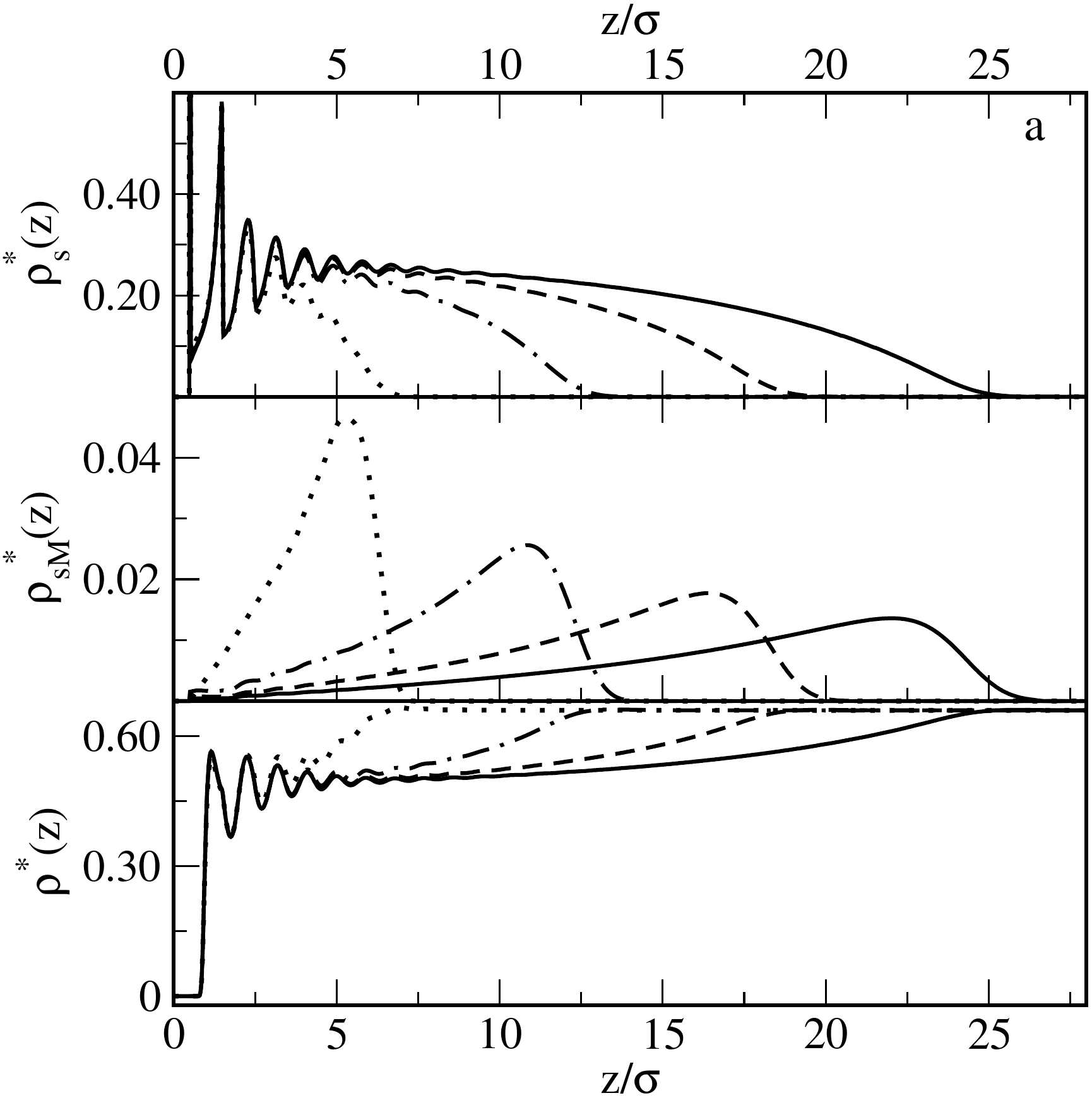}%
\hfill%
\includegraphics[width=0.48\textwidth]{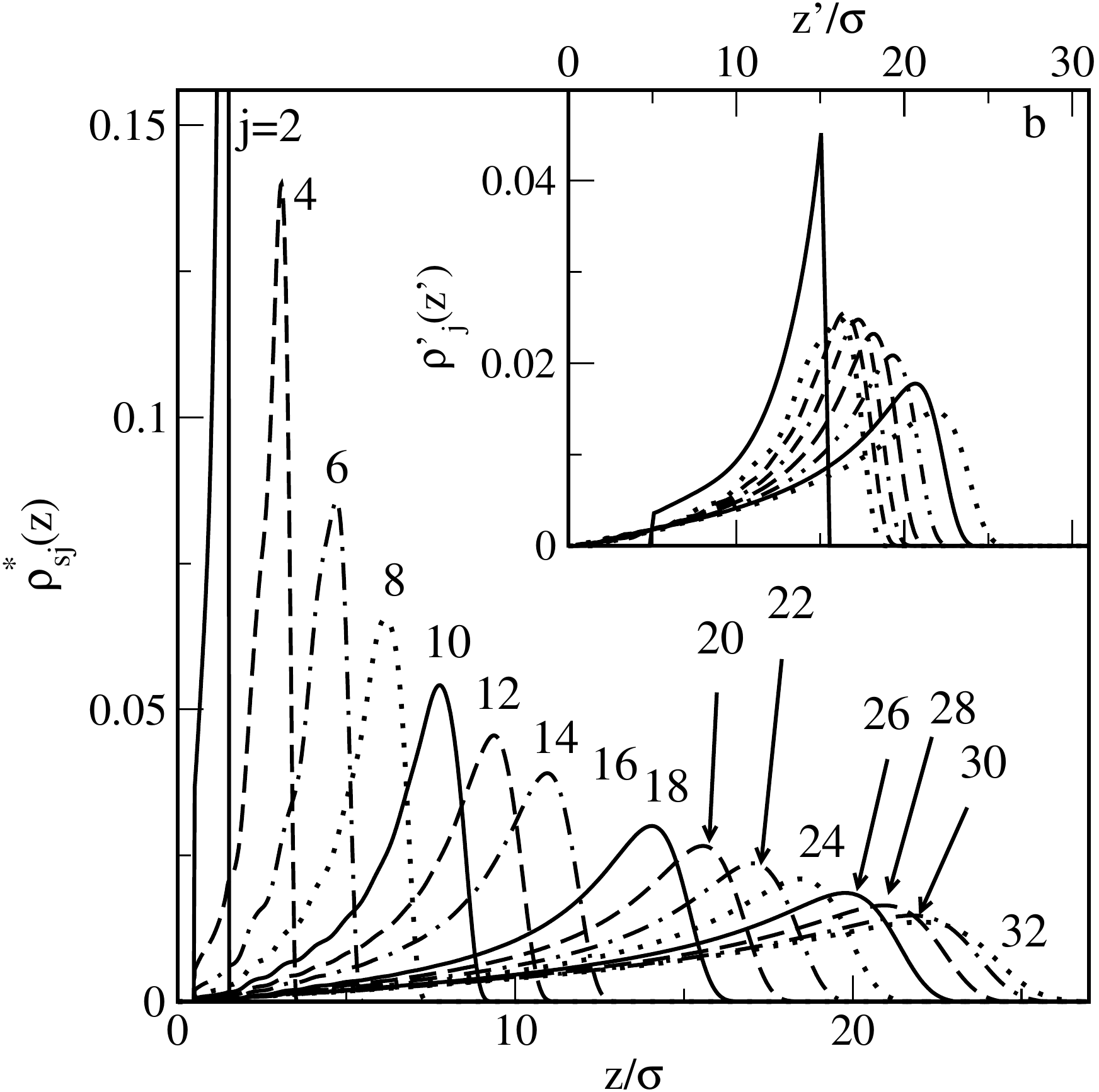}%
\\%
\parbox[t]{0.48\textwidth}{%
\centerline{(a)}%
}%
\hfill%
\parbox[t]{0.48\textwidth}{%
\centerline{(b)}%
}%
\caption{Part (a). Total segment density profiles (upper panel), the density profiles of the last segment (middle panel) and the
fluid density profiles for the chains of different length. Dotted line is for $M=8$,
dash-dotted line -- for $M=16$, dashed line is for $M=24$ and
solid line is for $M=32$.
Part (b) shows the density profiles of individual segments, $\rho_{\mathrm{s}j}^*(z)$ for the values of $j$ given
in the figure. Here the chain length is $M=32$. The inset to part (b) shows the scaled segment density profiles, as described in
the text. The consecutive curves from left to right are for $j=2$, 6, 10, 14, 18, 22, 26 and 30.
The segment-segment interactions are of hard-sphere type,
$\rho_{\mathrm{c}}^*=0.15$, $T^*=1$ and $\varepsilon_{\mathrm{fs}}^*=1$. }
\protect
\label{fig:2}
\end{figure}
\begin{figure}[ht]
\includegraphics[width=0.48\textwidth]{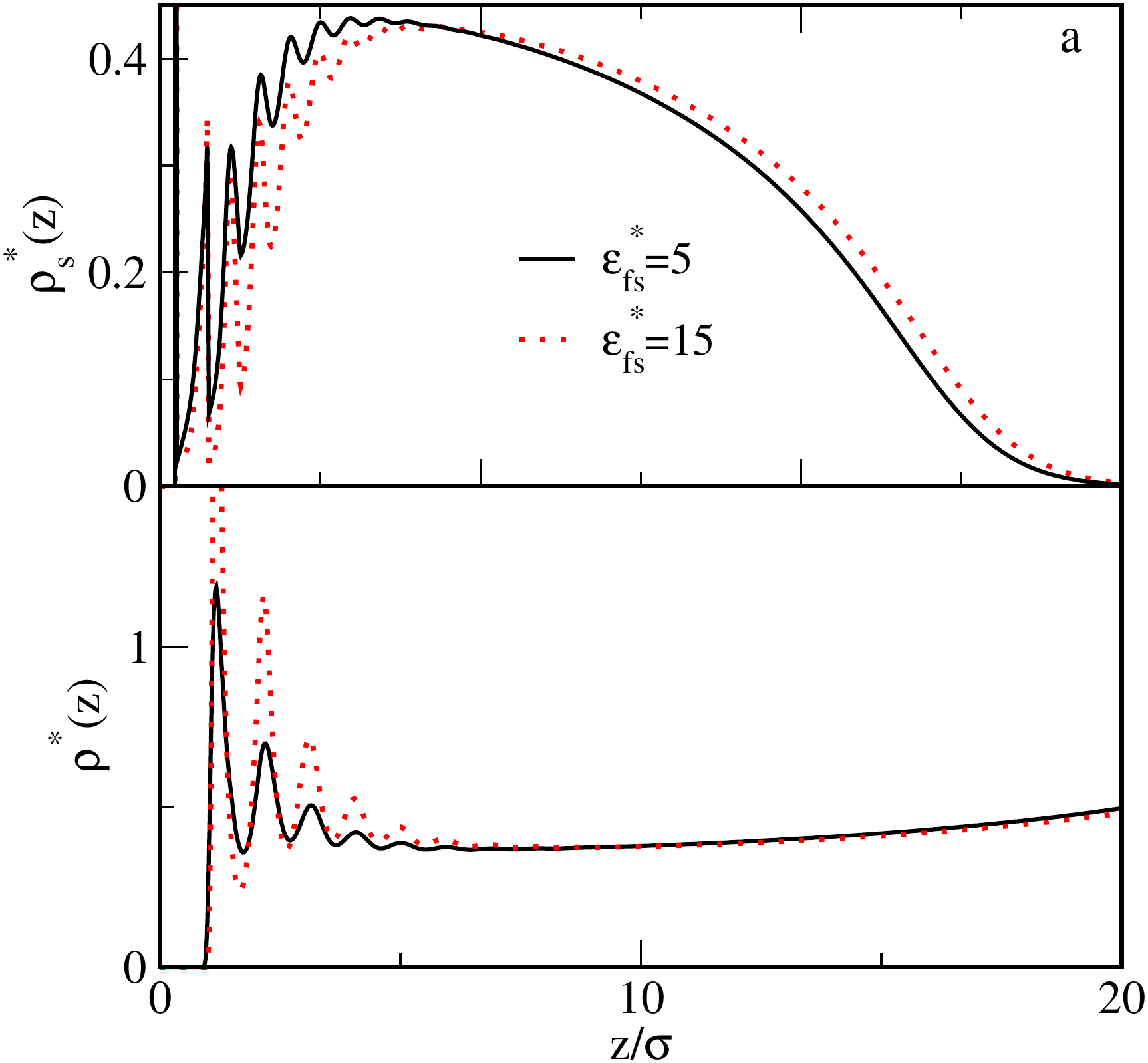}%
\hfill%
\includegraphics[width=0.48\textwidth]{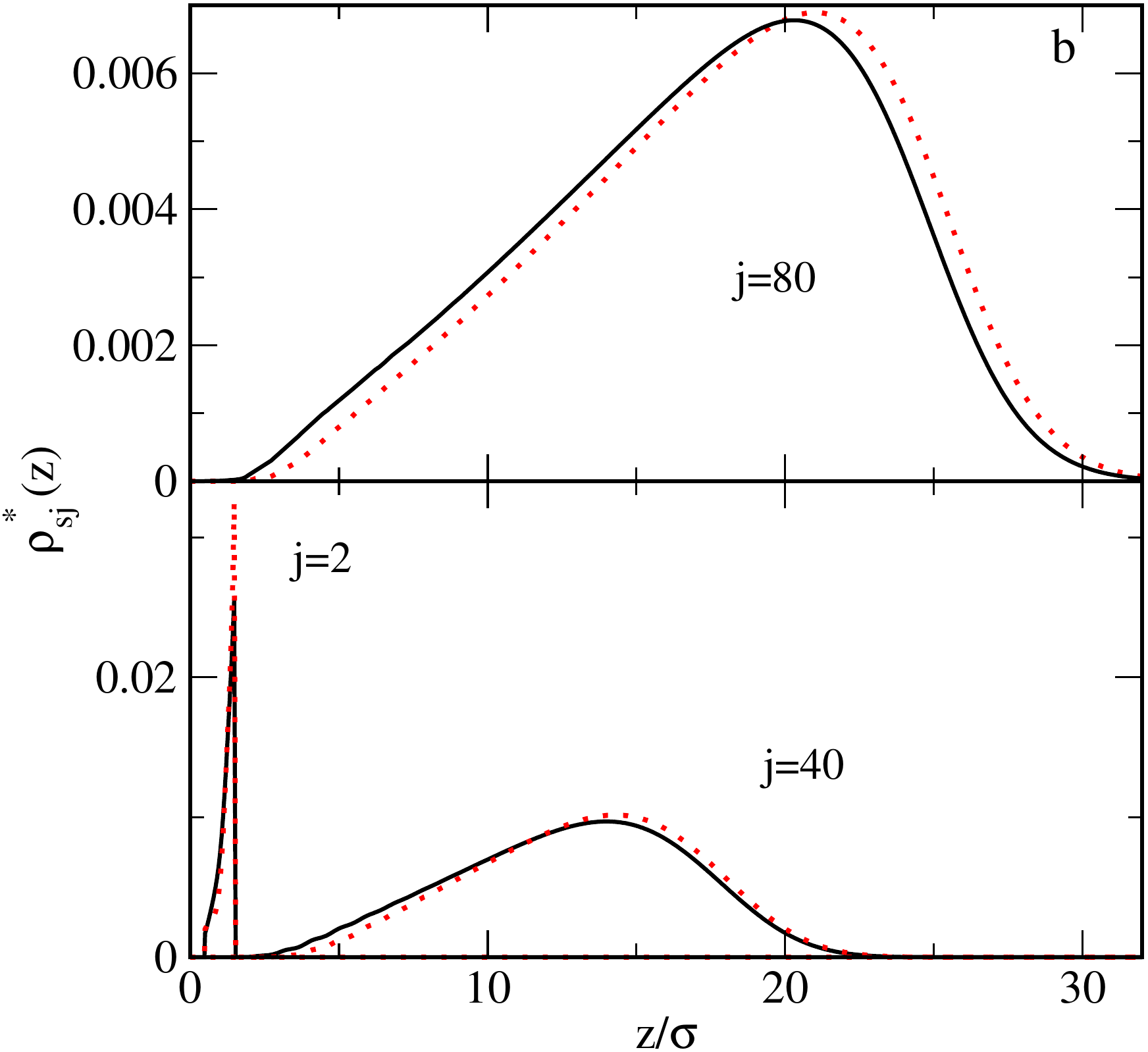}%
\\%
\parbox[t]{0.48\textwidth}{%
\centerline{(a)}%
}%
\hfill%
\parbox[t]{0.48\textwidth}{%
\centerline{(b)}%
}%
\caption{Part (a). Total segment density profiles  (upper panel) and fluid density profiles (lower panel)
of $M=80$-mers  for $\varepsilon_{\mathrm{fs}}^*=5$ (solid lines)
and $\varepsilon_{\mathrm{fs}}^*=15$ (dotted lines). In part (b) we display the profiles of the last segment (upper panel) and
of the middle segment ($j=40$) and of the second segment, $j=2$ (the latter are divided by 10). The values of the remaining parameters are
$R_{\mathrm{c}}^*=0.1$, $T^*=1$ and $\varepsilon_{\mathrm{cc}}=\varepsilon$.
 }
\protect
\label{fig:3}
\end{figure}

We have monitored the changes in the structures of the brush and of the fluid
with the changes of the parameters of the model
by inspecting the density profiles.  Examples of the obtained results are shown in figures~\ref{fig:2}
and~\ref{fig:3}. Figure~\ref{fig:2}~(a) shows how the total density profiles (the upper panel), the profiles of
the last segment (the middle panel) and the fluid density profiles (the lower panel) change with
the length of tethered chains. We see that if the length of the chains increases, the initial
parts (i.e. the part close to the wall) of the total segment density profiles and of the fluid density profiles
remain almost the same as for shortened chains.  An increase in the number of segments $M$ causes
expansion of the region where the segments and the fluid molecules exhibit a
layered structure. Instantaneously, the ``tails'' of the profiles become more diffuse for longer
chains. In the middle part of  figure~\ref{fig:2}~(a) we also see
that even for long chains there is nonzero probability of finding the last segments in the
vicinity of the wall. Such a shape of the segment density profiles suggest coiling of the chains which
can even lead to the formation of loops with both terminating segments of a chain located at the wall.

\newpage
Figure~\ref{fig:2}~(b) displays the profiles of individual segments with even numbers for tethered 32-mers.
With an increase of the segment number, the profiles become more and more diffuse.
According to classical self-consistent field theory, the re-scaled distributions of the segments
$\rho'_j(z')=\rho_{\mathrm{s}j}^*(z)\sin[\pi j/2M]$ should be an universal  function of $z'=z/ \sin[\pi j/2M]$, i.e.,
the plot of $\rho'_j$ versus $z'$  should be independent of the segment index $j$ ~\cite{58,59,60}.
The inset to figure~\ref{fig:2}~(b) provides a test
whether the above scaling can be also applied to the DFT results. Evidently,
the self-consistent field scaling fails when applied to the density functional results of this work.

Figure~\ref{fig:3}~(a) illustrates the change of the total segment density profile
 of tethered chains and of the profile of the fluid (lower and upper
panels, respectively), while figure~\ref{fig:3}~(b) shows the changes of the last
(lower panel) and of the middle (upper panel) with
an increase of the fluid-solid attraction. We see that when $\varepsilon_{\mathrm{fs}}^*$ increases, the
total segment density profile is pushed away from the solid surface by the adsorbing fluid.
This effect is more pronounced for the last segment than for the initial one of middle segments of the chain
(figure~\ref{fig:3}~(b)).

\subsection{The role of the surface density of the chains}

\begin{figure}[ht]
\hspace{3mm}
\includegraphics[width=0.45\textwidth]{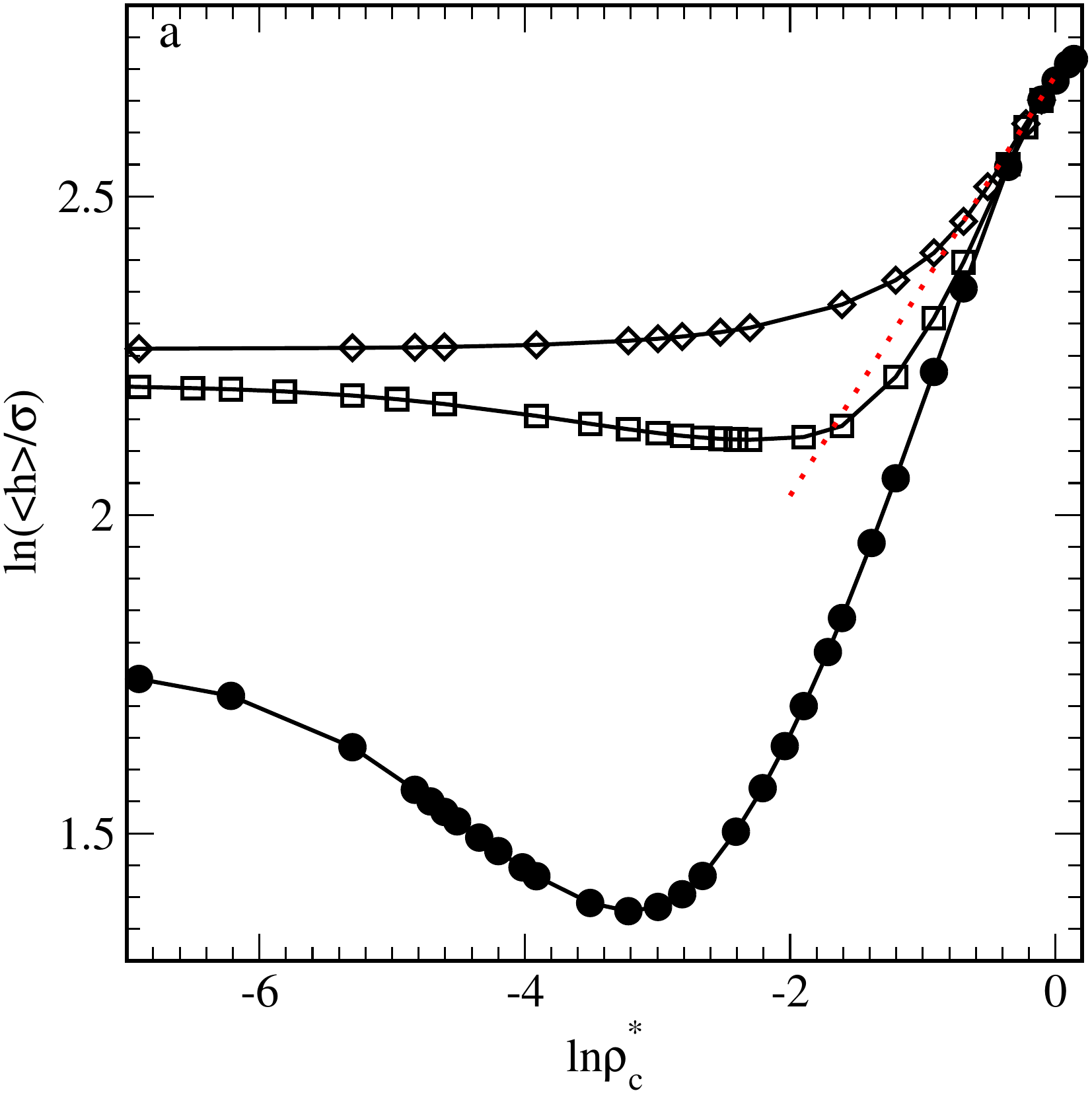}%
\hfill%
\includegraphics[width=0.45\textwidth]{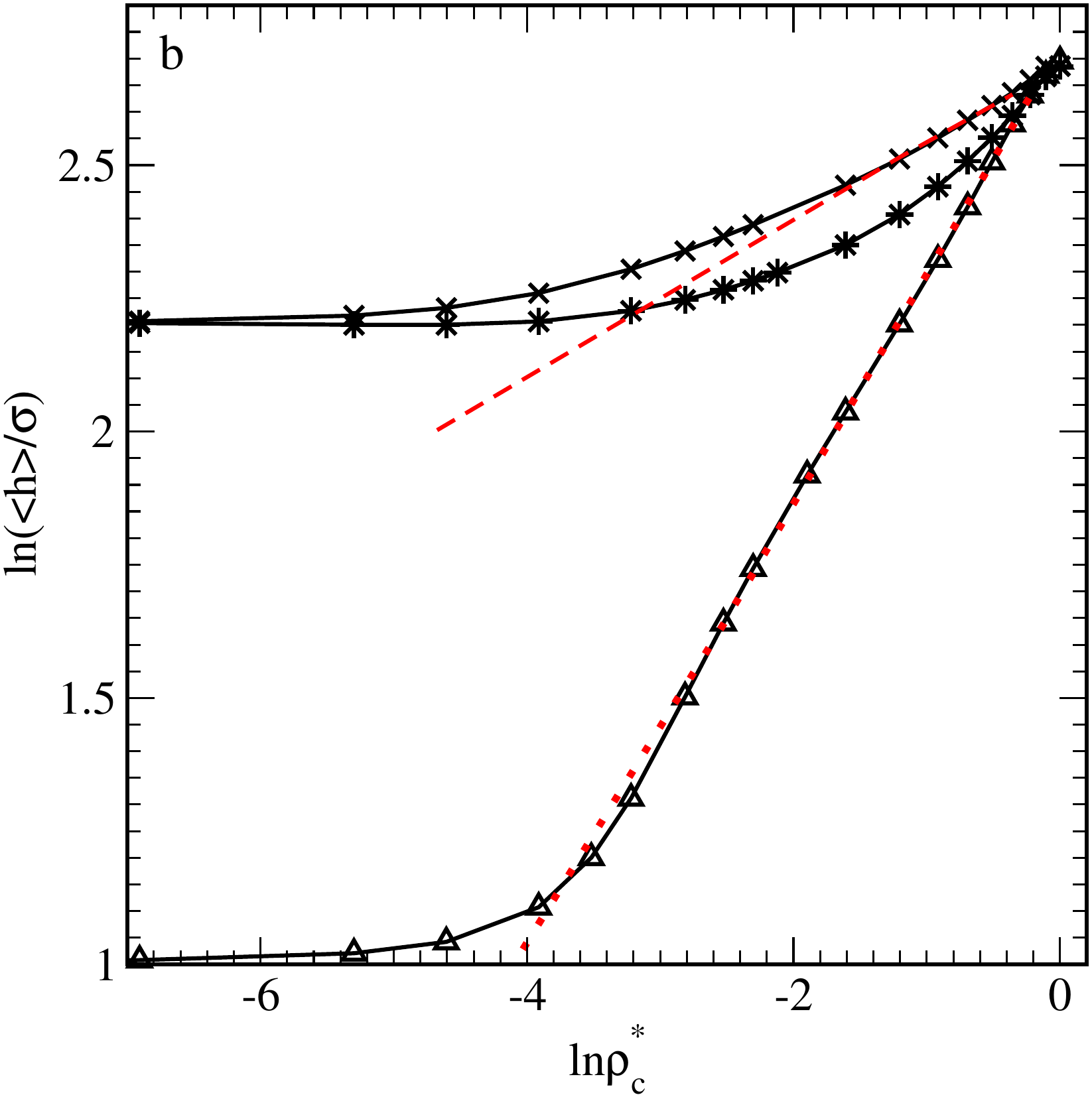}%
\hspace{3mm}
\\%
\parbox[t]{0.48\textwidth}{%
\centerline{(a)}%
}%
\hfill%
\parbox[t]{0.48\textwidth}{%
\centerline{(b)}%
}%
\caption{The dependence of $\langle h \rangle /\sigma$ on the surface density of tethered 16-mers. The nomenclature of the lines
in part (a) is as follows:  $\varepsilon_{\mathrm{fs}}^*=15$ and  $\rho_{\mathrm{b}}^*=0.66$ -- line with open diamonds;
 $\varepsilon_{\mathrm{fs}}^*=1$ and  $\rho_{\mathrm{b}}^*=0.66$ -- line with open squares. The line with solid
circles was obtained for the brush in a vacuum,  $\rho_{\mathrm{b}}^*=0$. In all cases
 $\varepsilon_{\mathrm{cc}}=\varepsilon$ and $T^*=1$. Dotted straight line has the slope of 1/3 and fits all
the results in the limit of high $\rho_{\mathrm{c}}^*$,  $\rho_{\mathrm{c}}^*\in[0.5,0.9]$.  In part (b) the consecutive
lines were obtained for the following parameters:  $\varepsilon_{\mathrm{cc}}=0.6\varepsilon$ and $T^*=1$ --
line with stars;  $\varepsilon_{\mathrm{cc}}=0$ (hard-sphere segment-segment interactions) and $T^*=1$ --
line decorated with crosses; $\varepsilon_{\mathrm{cc}}=\varepsilon$ and $T^*=5$ -- line decorated
with triangles. In all cases  $\rho_{\mathrm{b}}^*=0.66$ and  $\varepsilon_{\mathrm{fs}}^*=1$.
 The dotted and dashed straight lines denote the approximations of the log-log plots, as
described in the text. }
\protect
\label{fig:4}
\end{figure}

\looseness=-1The next problem we want to study in detail is how the surface density of the pinned chains effects
the height of tethered layer.
We begin our discussion with the results for
 rather short chains built of 16-mers (the solids modified with the
chains of a comparable length are widely used as column fillings in chromatography).
In figure~\ref{fig:4} we demonstrate the
effect of some selected parameters of the model  on the relationship of  $\langle h \rangle /\sigma$ upon $\rho_{\mathrm{c}}^*$.
Three curves displayed in part (a) have been calculated assuming that the segment-segment interactions are
the same as the segment fluid and fluid-fluid interactions, $\varepsilon_{\mathrm{cc}}=\varepsilon$.
The temperature was kept constant and equal to $T^*=1$.
The curve  decorated with open squares is for the solid that weakly attracts fluid molecules,
$\varepsilon_{\mathrm{fs}}^*=1$, while the curve decorated with open diamonds is for strongly
attracting solid surface, $\varepsilon_{\mathrm{fs}}^*=15$.  These two curves were evaluated assuming
that the fluid density was $\rho_{\mathrm{b}}^*=0.66$. However, we also present here the results
 for tethered chains in a vacuum (curve decorated with solid circles). Part (b) illustrates
the effects of weakening of the segment segment interactions and of the temperature
on $\langle h \rangle /\sigma$.

For very low surface densities of tethered chains, the brush height remains almost constant.
This is not surprising, because under such conditions a tethered chain does not almost
``feel'' the presence of other chains and assumes the configuration that
is only effected by the presence of  solvent molecules.  The accumulation of
solvent molecules in the vicinity of the solid surface leads to an
increase of $\langle h \rangle /\sigma$. Of course, the last statement is true if the solvent-fluid
attraction is similar (in our case identical) to the fluid-fluid attraction.
The behavior of the brush in ``bad'' solvent can be different~\cite{38b}, but this problem
requires additional studies.

With further increase of $\rho_{\mathrm{c}}^*$ and when the attractive fluid-solid potential
is weak, there develops a minimum on the curve of $\langle h \rangle /\sigma$ vs. $\rho_{\mathrm{c}}^*$
 (see the line with open squares in figure~\ref{fig:4}~(a)). In this case,
the minimum is
located at $\rho_{\mathrm{c}}^*\approx 0.1$. Similarly, a minimum on the curve of
 $\langle h \rangle /\sigma$ vs. $\rho_{\mathrm{c}}^*$ is observed if the brush is in contact
with a vacuum, but now it is located at much lower surface brush density,
 $\rho_{\mathrm{c}}^*\approx 0.04$.
 For a strongly
attractive wall, this minimum disappears (solid line with crosses in figure~\ref{fig:4}~(a)).
 Also, the minimum vanishes  when the
segment-segment attraction is lowered or when the temperature is raised, see figure~\ref{fig:4}~(b).
Evidently, this phenomenon is caused by the segment-segment attractive interactions.
At low temperatures, the attraction between segments causes coiling of the chains near the
wall. However, in the case of the wall strongly attracting fluid molecules, the coiling
is inhibited by the fluid molecules accumulated at the wall.

In general,
two characteristic regions on the curves
$\langle h \rangle /\sigma$ vs. $\rho_{\mathrm{c}}^*$ in figures~\ref{fig:4}~(a) and \ref{fig:4}~(b) can be distinguished: the region of almost
constant values of $\langle h \rangle /\sigma$ at
low surface densities of tethered chains and the region where the logarithm of the height of the brush
changes with $\ln\rho_{\mathrm{c}}^*$ almost linearly. Such a behavior was also observed in computer simulations~\cite{32a,52}.

Classical scaling theories predict that the height of the brush scales with the
surface brush density as
\begin{equation}
\langle h \rangle \propto (\rho_{\mathrm{c}})^\gamma
\label{eq:sc2}
\end{equation}
 with $\gamma=1/3$, as predicted by the
self-consistent mean-filed theory (see, e.g.,~\cite{32a}), or with $\gamma \approx  0.35$, as predicted
by the Alexander--de Gennes approach (see, e.g.,~\cite{52}. The dotted line plotted in figure~\ref{fig:4}~(a) has the slope of 1/3
and it well fits all the data for $0.5<\rho_{\mathrm{c}}^*<0.9$.  However, for the values of $\rho_{\mathrm{c}}^*$ higher than 0.9
the changes of $\langle h \rangle /\sigma$ are much smaller than predicted by equation~(\ref{eq:sc2}). Of course, for
a very dense brush, its height attains a maximum corresponding to fully stretched
chains.

 At high temperature, the linearity range of the log-log plot is much wider, cf. dotted line
in figure~\ref{fig:4}~(b). This line well approximates  the $\langle h(\rho_{\mathrm{c}}^*)/\sigma \rangle$
curve evaluated at $T^*=5$ (the line with triangles), but now the exponent
$\gamma\approx 0.43$ is significantly higher than predicted by the theory of Alexander and
de Gennes~\cite{21,22,22a}.   Similarly, when the segment-segment interactions are
of hard-sphere type, there exist a range of surface densities $ \rho_{\mathrm{c}}^*$, within
which the log-log plot is linear (dashed line in figure~\ref{fig:4}~(b)), but now the slope is much lower,
$\gamma \approx 0.14$, than predicted by the classical theories. Only when
$\varepsilon_{\mathrm{cc}}=0.6\varepsilon$ (the curve with stars in figure~\ref{fig:4}~(b)),
the straight line approximating the data
in a relatively wide range of  $\rho_{\mathrm{c}}^*$ values
has the slope close to 1/3 (the relevant line has been omitted in figure~\ref{fig:4}~(b)).

\begin{figure}[ht]
\includegraphics[width=0.32\textwidth]{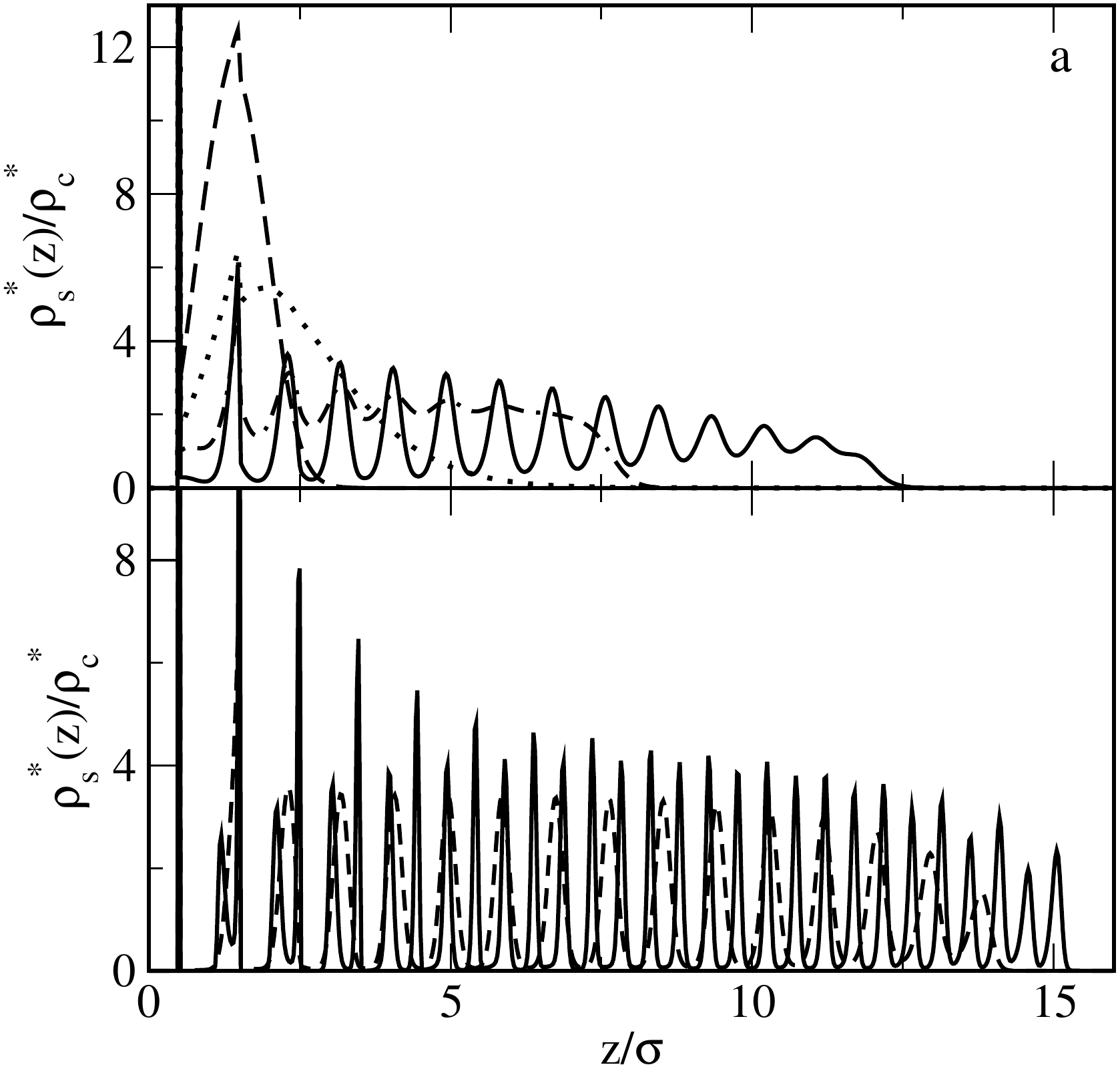}%
\hfill%
\includegraphics[width=0.32\textwidth]{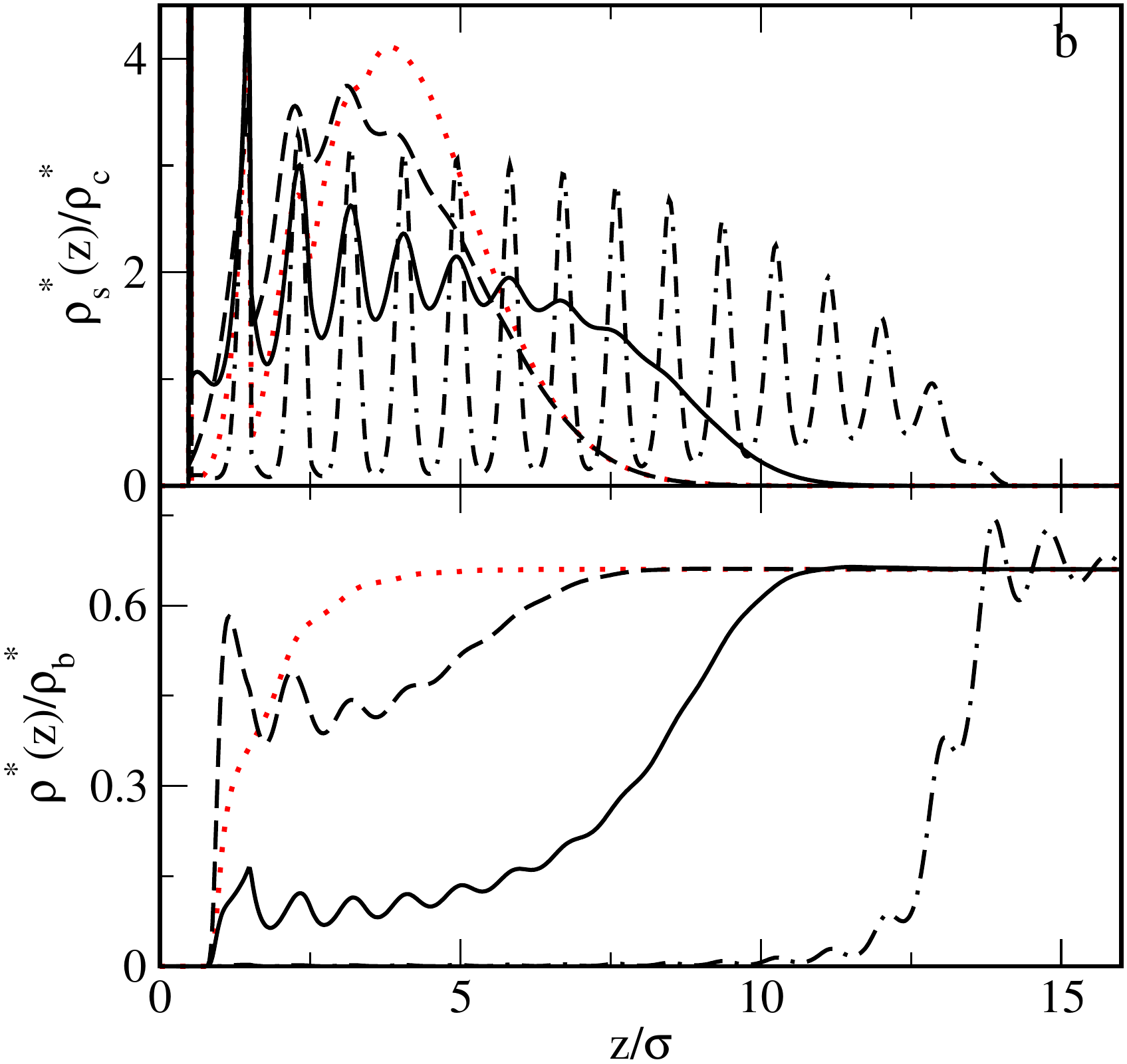}%
\hfill%
\includegraphics[width=0.32\textwidth]{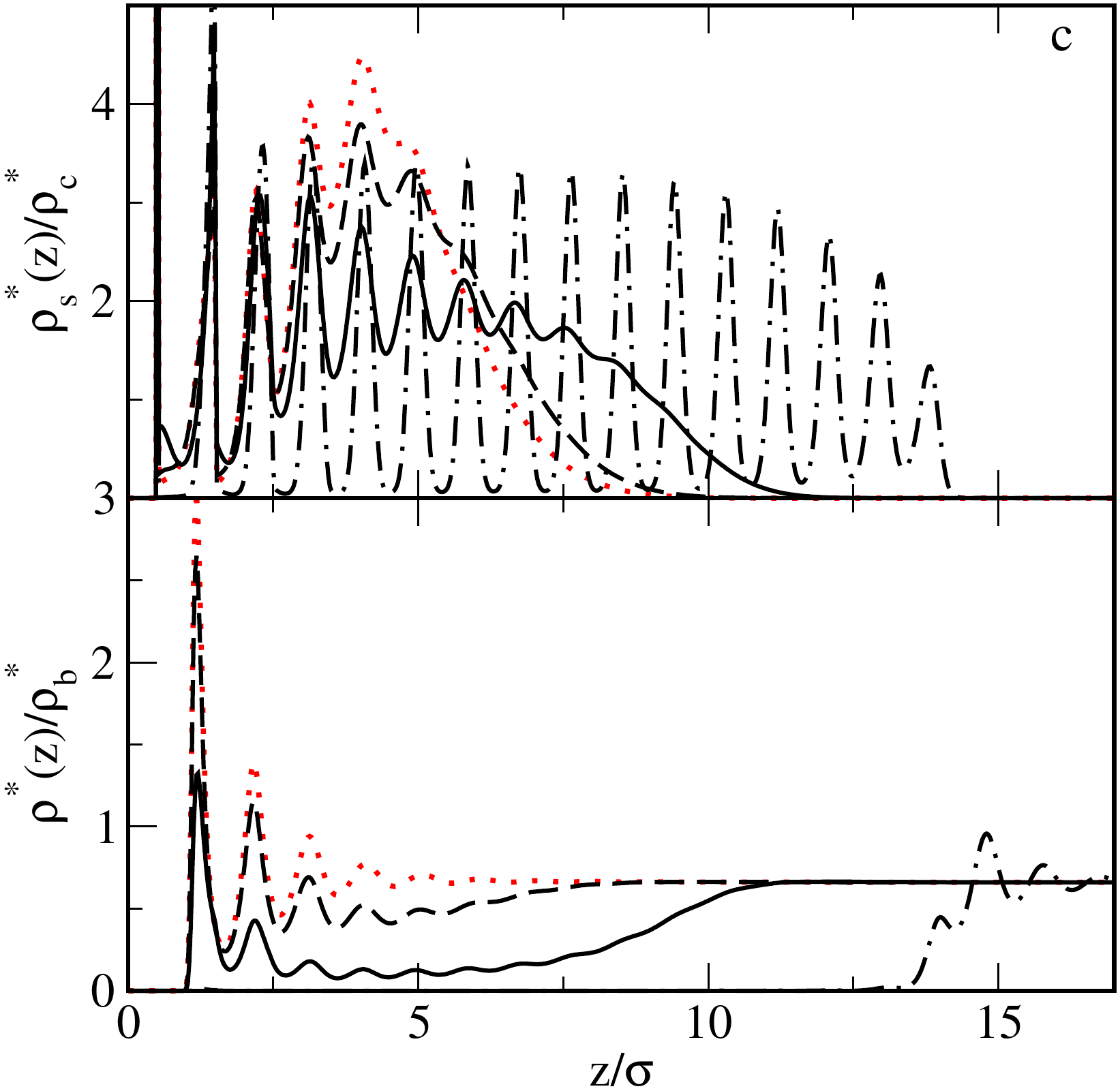}%
\\%
\parbox[t]{0.32\textwidth}{%
\centerline{(a)}%
}%
\hfill%
\parbox[t]{0.32\textwidth}{%
\centerline{(b)}%
}%
\hfill%
\parbox[t]{0.32\textwidth}{%
\centerline{(c)}%
}%
\caption{Examples of the total segment density profiles  of 16-mers and the profiles of fluid calculated along
the curves displayed in figure~4~(a). Part (a) is for the brush in contact with a vacuum (line with
solid circles in figure~4). Upper panel shows the results for $\rho_{\mathrm{c}}^*=2\cdot 10^{-6}$~-- dotted line,
0.04 (this surface density corresponds to the minimum of  $\langle h \rangle /\sigma$)~-- dashed line, 0.4~-- dash-dotted linear and
0.7 -- solid line. Lower panel is for $\rho_{\mathrm{c}}^*=0.9$~-- dashed line and for 1.15~-- solid line.
Part (b) was calculated along the line with open squares in figure~4~(a). Upper panel displays the
total segment density profiles and lower panel displays the profiles of fluid. The calculations are for
  $\rho_{\mathrm{c}}^*=2\cdot 10^{-6}$~-- dotted lines, 0.1  (this surface density corresponds to the minimum of  $\langle h \rangle /\sigma$)~-- dashed line, 0.4~-- solid line and 0.9~-- dash-dotted line.
 Part (c) was calculated along the line with open squares in figure~4~(a). The abbreviation of the
lines and the surface densities of tethered chains are the same as in part (b).
 }
\protect
\label{fig:5}
\end{figure}
Figure~\ref{fig:5} illustrates how the structure of the brush built of 16-mers in a vacuum changes with $ \rho_{\mathrm{c}}^*$. At
extremely low surface density, $ \rho_{\mathrm{c}}^*=10^{-6}$,  the brush extends to the distance around
$z/\sigma \approx 5.5$ and the profiles $\rho_{\mathrm{s}}^*(z)/\rho_{\mathrm{c}}^*$ remain almost unchanged up to
$\rho_{\mathrm{c}}^*\approx 5\cdot 10^{-5}$ (the relevant results have been omitted for the sake of brevity).  Then,
the coiling process starts due to attraction between the segments and the minimum height of the brush
is attained at $ \rho_{\mathrm{c}}^*\approx 0.04$. The profile corresponding
to the  brush of minimal height is shown by the dashed line in the upper panel of figure~\ref{fig:5}~(a). This
curve demonstrates that  the
formed structure  essentially comprises two layers at the wall.
Further increase of $ \rho_{\mathrm{c}}^*$ leads to the brush expansion. Initially,
the layer at the wall resembles a liquid-like film, but then a layered
structure develops. For  $ \rho_{\mathrm{c}}^*<0.9$, the number of layers is lower than
the number of the segments $M$, but at very high values of $ \rho_{\mathrm{c}}^*$, the
number of layers increases up to $2M$ (see the lower panel in figure~\ref{fig:5}~(a)).  The brush
layer tends to assume the solid-like
configuration~\cite{60a}. Unfortunately, one-dimensional density functional theory used in
this work fails in that region of the surface brush densities.

The presence of the fluid changes the structure of the brush, cf. figures~\ref{fig:5}~(b) and \ref{fig:5}~(c).
When $\rho_{\mathrm{c}}^*$ is not too high (say, $\rho_{\mathrm{c}}^*<0.4$, the fluid
itself forms layered structure, especially for higher values of $\varepsilon_{\mathrm{fs}}^*$.
The presence of a well-pronounced fluid layers in the vicinity
of the wall inhibits coiling of the brush and breaks the development of a minimum
on $\langle h(\rho_{\mathrm{c}}^*)/\sigma \rangle$ curve. The layers of the fluid  also cause an increase
of the brush height, in comparison with the brush in a vacuum.
However, for higher surface densities of tethered chains, the fluid
molecules are no longer capable of penetrating the interior of
the brush. Consequently, for $\rho_{\mathrm{c}}^*>0.5$, the heights of brushes formed
in the systems with and without the presence of the fluid,
 become similar and when  $\rho_{\mathrm{c}}^*$  still  increases the
values of $\langle h(\rho_{\mathrm{c}}^*)\rangle/\sigma$ nearly flow together. The last statement is true
if the segment-segment, segment-fluid and fluid-fluid interactions in the
systems characterized by different values of  $\rho_{\mathrm{b}}^*$ and $\varepsilon_{\mathrm{fs}}^*$
are identical (or similar).  However,  when those interactions
are different, then for high values of   $\langle h(\rho_{\mathrm{c}}^*)/\sigma \rangle$, the differences
in the brush height in different systems remain significant, cf. figure~\ref{fig:4}~(b).

\begin{figure}[ht]
\begin{center}
\includegraphics[width=0.45\textwidth,clip]{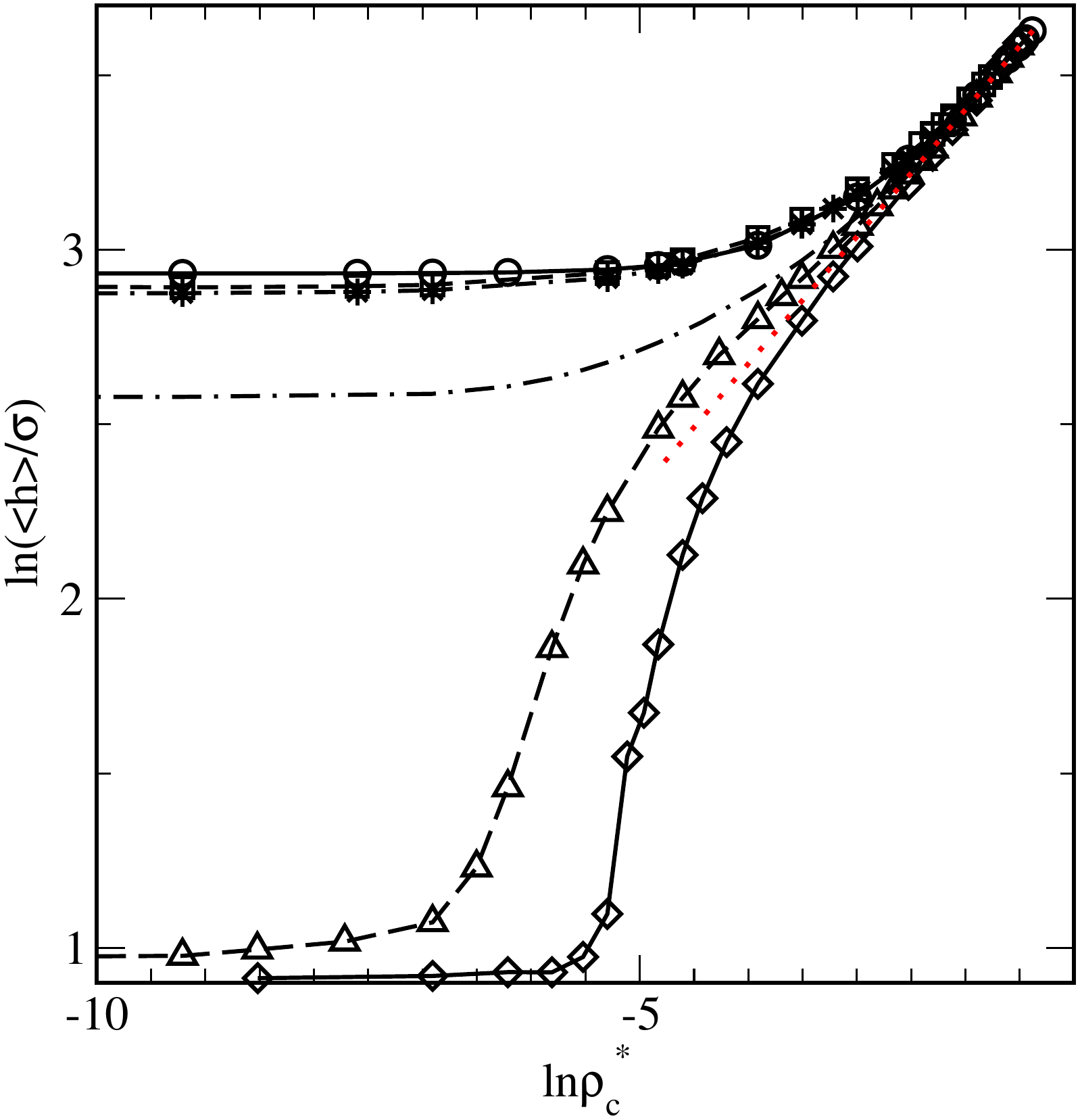}
\end{center}
\caption{The dependence of $\langle h \rangle /\sigma$ on the surface density of tethered 80-mers. The nomenclature of the lines
is as follows: $T^*=1$~-- solid line with open circles, $T^*=2$~-- dashed line with open squares,
$T^*=2.2$~-- double dash-dotted line with stars, $T=2.75$~-- dash-dotted line, $T^*=3$~-- dashed line with
triangles and $T^*=4$~-- solid line with diamonds. The calculations are for
$\varepsilon_{\mathrm{fs}}^*=12$, and $\varepsilon_{\mathrm{cc}}=\varepsilon$. Dotted straight line has the slope 0.35.
 }
\protect
\label{fig:6}
\end{figure}
The results presented in figures~\ref{fig:4} and \ref{fig:5} have been computed for rather short chains
composed of 16-mers. We now consider the case of 80-mers. Figure~\ref{fig:6} shows
the functions $\langle h(\rho_{\mathrm{c}}^*)/\sigma \rangle$ evaluated at different temperatures for the
system characterized by $\varepsilon_{\mathrm{cc}}=\varepsilon$ and $\varepsilon_{\mathrm{fs}}^*=12$.
Similarly to the case of $M=16$ at low surface densities of grafted chains, we observe here that
the heights of the brushes remain almost constant. For high values of
$\rho_{\mathrm{c}}^*$, the curves
 at different temperatures almost coincide and all of them can be approximated
by a straight line with the slope of 0.35, in accordance with the theory by Alexander and
de Gennes. However, we observe here an unexpected effect. Namely, with an increase
of the temperature, the average height of the low surface density brush  decreases.
This decrease is small when the temperature rises from $T^*=1$ up to $T=2.2$,
but then the changes become larger, especially within the temperature range of $[2.75,3]$.
We carefully inspected that region of temperatures and found that the occurring changes are
fast, but continuous, i.e., the plot of the thermodynamic potential $Y$  vs $T^*$  at a constant
$\rho^*_{\mathrm{c}}$ does not exhibit any discontinuities.

\begin{figure}[ht]
\begin{center}
\includegraphics[width=0.6\textwidth,clip]{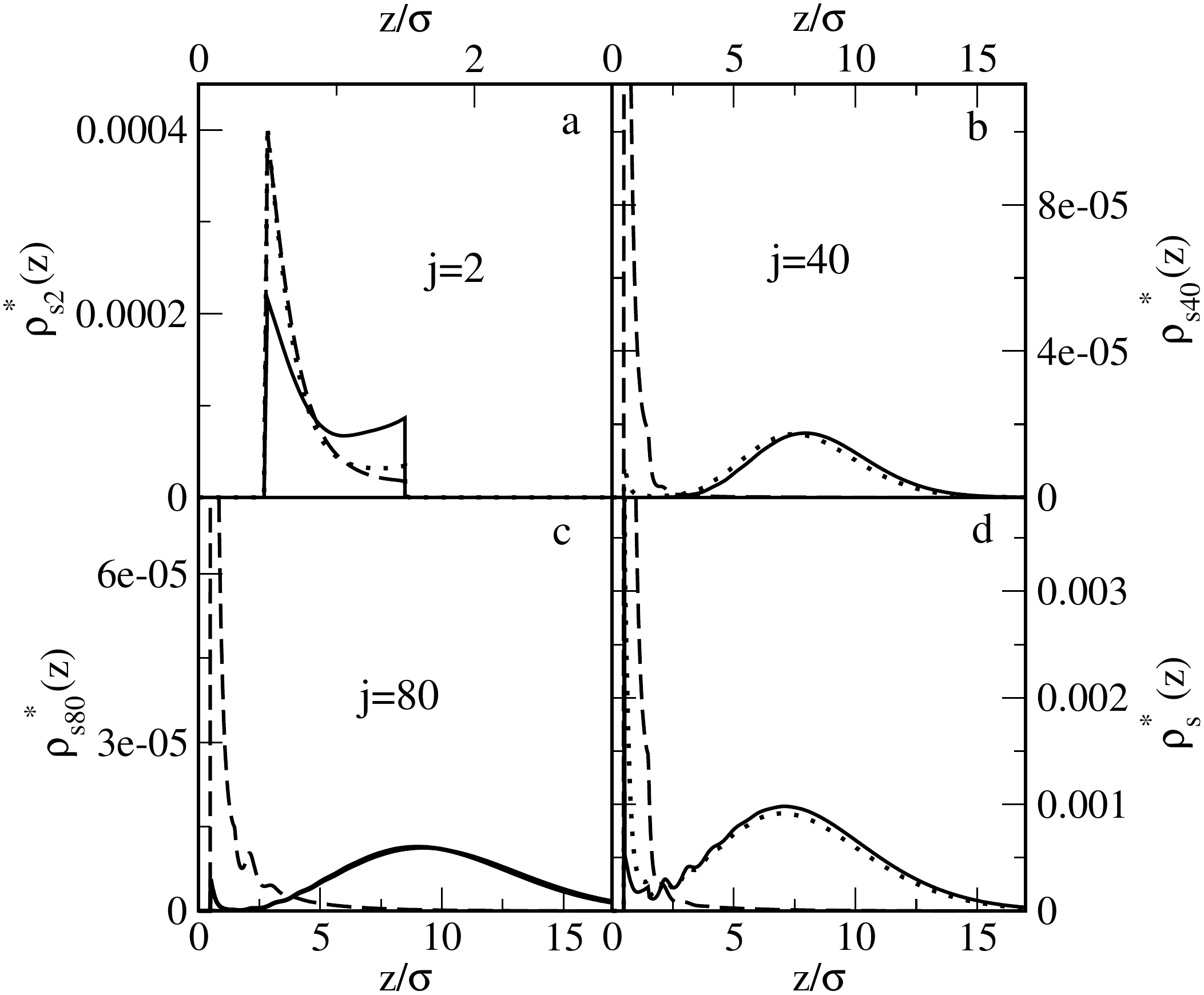}
\end{center}
\caption{Segment density profiles $\rho_{\mathrm{s}j}(z)$  for $j=2$ (a), $j=40$ (b),
$j=80$ (c) and the total segment density profiles (d).
The calculations were carried out for the systems from figure~\ref{fig:6} at $\rho_{\mathrm{c}}^*=0.0001$ and
at  $T^*=2.2$ (solid lines) and at $T^*=3$ (dashed lines).  Dotted lines show the
results evaluated at the same surface density of the chains and at $T^*=3$, assuming that
x $\varepsilon_{\mathrm{fs}}^*=18$.  }
\protect
\label{fig:7}
\end{figure}
In order to explain the observed temperature behavior we have inspected the density profiles.
In figure~\ref{fig:7} we display the segment density profiles of the segment $j=2$, 40 and  80
 and the total segment density profiles.
The calculations were carried out for the system presented in figure~\ref{fig:6} at $\rho_{\mathrm{c}}^*=0.0001$ and
at  $T^*=2.2$ (solid lines) and at $T^*=3$ (dashed lines). Additionally we have displayed here the
results evaluated at the same surface density of the chains and at $T^*=3$, but assuming that the
fluid-solid energy is  higher, $\varepsilon_{\mathrm{fs}}^*=18$ (dotted lines).
 The temperature $T^*=3$ is well below the
bulk critical temperatures of the fluid comprising 80-mers, cf.~\cite{61} and thus
the chains exhibit a tendency to condensate, even within a single chain. At lower temperature
$T^*=2.2$  the adsorption of the fluid at the wall is stronger than at $T^*=3$. The accumulation
of the fluid particles at the wall causes the situation wherein  not all the segments of the chain ``condensate'', i.e., the initial segments do not join the ``cloud'' formed by the inner and outer segments and the
tethered chains assume a characteristic mushroom-like shape. Indeed, the profiles
for $j=40$ and for $j=80$ displayed in figure~\ref{fig:7} lie very close to each other and the total
segment density profile exhibits a depletion for $0.5<z<2.5$ (this depletion corresponds to
 the  mushroom's   ``pedicle''). When the temperature is raised up to $T^*=3$, a part of the
fluid molecules is removed from the surface region, making more room for the segments
of the chains. Consequently, the chains  are capable of assuming the
configuration almost parallel to the wall and then they ``condensate'' into a thin
(in comparison with the mushroom configuration) film covering the wall (pancake configuration) . However, when
the wall attracts the fluid particles more strongly, $\varepsilon_{\mathrm{fs}}^*=18$, the
formation of a thin film of chains at the wall is inhibited by the presence of the fluid
particles and the brush assumes a mushroom structure, even at $T^*=3$. The phenomenon
of the change of the brush structure is connected with an interplay between adsorption
of the fluid (the formation of layered structure by the fluid molecules) and the tendency
of chains to condensate. In the case of systems under study, this change is a continuous process,
but we cannot exclude a situation when this transformation will occur as a first-order transition.

The relations describing the changes in the brush height with the changes of the number of segments
and with the grafting density, known as scaling relations~\cite{21,22,22a,23,24,32a,38,52,56} were
developed using theoretical description based on coarse-grained models. Therefore, it is not surprising
that these relations agree well with the coarse-grained simulations (e.g., Monte Carlo simulations that are
based on the self-avoiding random walk model of chains), However, when microscopic modelling is
applied in computer simulations, several deviations between computer simulation data and
scaling theory arise, see, for example~\cite{32a}.

In our calculations we observe that when the strength of
the segment-segment interactions decreases, the scaling relations are approximately satisfied within
the region of stretched chains. However, more importantly, we have found two unexpected effects.
The first one is a possibility of the existence of a minimum on the curve describing the brush height
versus the grafting density, while the second one is a decrease of the brush height with an increase of temperature
(or an increase of the brush height with a decrease of temperature).
Both effects have been observed at rather low grafting densities. The second effect results, in our opinion, from
temperature dependence of adsorption of solvent molecules. In general, the effect of the adsorption
of solvent on the surface on the brush height has been omitted in previous studies. To our
best knowledge, there are no relevant computer simulation data that would confirm the
existence of the two  unusual effects mentioned above. Such simulations require long-lasting calculations
and the question  whether our observations are artifacts of the theory or they do really exist is currently
under study in our laboratory.

\section*{Acknowledgements}
S.S. was supported by the Ministry of Science of Poland
under Grant No. N N204 151237. M.B., A.P. and O.P. acknowledge support from EC under Grant
No.~PIRSES 268498.

\ukrainianpart

\title{Зміни у структурі зв'язаних ланцюгових молекул згідно передбачень методу функціоналу густини}
\author{М. Боровко\refaddr{label1}, А. Патрикєєв\refaddr{label1},
О. Пізіо\refaddr{label2}, С. Соколовскі\refaddr{label1} }

\addresses{
\addr{label1} Відділ моделювання фізико-хімічних процесів, університет Марії Кюрі-Склодовської, \\ 20031 Люблін, Польща
\addr{label2} Інститут хімії УГАМ, 04360 Мехіко, Мексика
}

\makeukrtitle

\begin{abstract}
\tolerance=3000%
Ми використовуємо один із варіантів теорії функціоналу густини для вивчення змін у висоті зв'язаного шару ланцюгів, утворених з'єднаними сферичними сегментами, із змінними довжиною та поверхневою густиною ланцюгів. Для моделі, у якій взаємодії між сегментами та молекулами розчинника є такими самими як між молекулами розчинника, нами виявлено два ефекти, які не спостерігалися у попередніх дослідженнях. За певних умов і при низьких поверхневих концентраціях ланцюгів висота зв'язаного шару може досягати мінімуму. Крім того, для деяких систем спостерігається що при збільшенні температури висота шару ланцюгів може зменшуватися.
\keywords щітка, адсорбція, теорія функціоналу густини, масштабування
\end{abstract}

\end{document}